# ARTICLE

# Hydride Ion Intercalation and Conduction in the Electride $Sr_3CrN_3$

Xu Miaoting,[a] Cuicui Wang,[a] Benjamin J. Morgan[b] and Lee A. Burton*[a]

The electride $Sr_3CrN_3$ has a one-dimensional channel of electron density, which is a rare feature that offers great potential for fast ion conduction. Using density functional theory, we find that $Sr_3CrN_3$ is an excellent hydride conductor within this channel, with a diffusion barrier as low as 0.30 eV and an estimated diffusion coefficient of $5.37×10^{-8}$ cm$^2$/s. This diffusion barrier is lower than those reported for the best hydride conductors to date. We also show the most-stable amount of hydride in the host material under standard conditions and the corresponding change in electronic structure from metal to wide-gap insulator. Our results highlight the potential offered by 1D electride materials for ion-transport applications such as energy storage or gas separation.

## Introduction

The global emphasis on restricting carbon emissions continues to increase the demand for clean technologies.[1] For example, renewable energies are predicted to phase-out coal and gas power, while electric vehicles are projected to replace those with internal combustion engines.[2] Both of these advancements require continuing advances in electrochemical energy storage technologies, such as lithium-ion batteries or fuel cells.[3]

Solid-state ion-conducting materials have been long-studied for their potential use in energy storage systems.[4] Recent years have seen particular advances in the development of highly-conducting lithium ion solid electrolytes with potential application in all-solid-state lithium-ion batteries.[5–8] The limited global availability of lithium, however, means that lithium-ion batteries are expected to meet only part of the projected future energy-storage needs.[9] This motivates the continuing search for alternate materials, including solid-state electrolytes, that could be used in non-lithium energy-storage devices.

Hydride ions have small ionic radii, large electronic polarizability and a high standard redox potential for $H_2/H^-$ (−2.3 V).[10] These characteristics make $H^-$ ions promising for applications in next-generation electrochemical energy storage with high voltage and high energy density. Furthermore, hydrogen is highly abundant, globally available and low cost. To date, materials with high ion conduction for $H^-$ have been reported but these were only achieved at relatively high temperatures.[12] Thus, there is still ample scope for improvement in this area.

Electrides are class of rare ionic compounds that possess free electrons localized within cavities in the host structure, with these electrons acting as anions. To date, only a few inorganic electrides have been experimentally identified,[13] but they are typically classified according to dimensionality of their free electron density. For example, if the excess electron density occupies cavities or pores the electrides are classified as 0D (e.g., $Ca_{12}Al_{14}O_{32}$[14]), if the density is in a one-dimensional channel they are 1D (e.g. $Y_5Si_3$[15]) or if the density is continuous in a plane they are 2D (e.g. $Ca_2N$[16]). The anionic electrons of electrides have been shown to effectively interact with external hydrogen, leading to excellent hydride absorption and desorption properties,[17,18] even in 1D electrides.[15]

In this work we consider $Sr_3CrN_3$, a recently identified 1D electride, as a hydride conductor.[19] $Sr_3CrN_3$ can be described with the formal oxidation states $Sr^{2+}{}_3Cr^{4+}N^{3-}{}_3{:}e^-$ and the excess electrons aggregating in one-dimensional channels (see **Figure 1**).[20] Here, we explore the optimal diffusion path of hydride ions in the channel using the climbing-image nudged-elastic-band (CI-NEB) method. We also investigate the hydrogen capacity by calculating insertion energies of hydride ions in the structure. Finally, we report the electronic nature of the host material before and after hydride intercalation, showing a shift from metallic to insulating behaviour. Overall, we find that $Sr_3CrN_3$ has a low capacity for hydrogen under standard conditions but the diffusion barrier of 0.3 eV within the 1D channels is significantly lower than barriers reported for best-in-class hydride ion conductors previously e.g. 0.52 eV for barium hydride.[12]

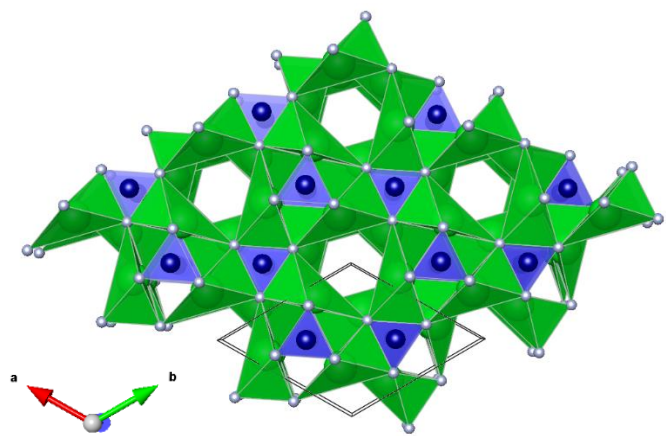

**Figure 1** The crystal structure of $Sr_3CrN_3$ with Sr, Cr and N atoms represented by green, blue and light grey spheres respectively. The unit cell is shown with black lines. The empty channels in c-direction are where the excess electron density resides.

[a.] International Centre for Quantum and Molecular Structures, Department of Physics, Shanghai University, Shanghai 200444, China.
[b.] Department of Chemistry, University of Bath, UK.

## Methods

Density function theory (DFT) calculations were performed using the Vienna Ab Initio Simulation Package (VASP),[21,22] with the Projector Augmented Wave (PAW) method for modelling core electrons.[23,24] The Perdew-Burke-Enzerhof (PBE) exchange-correlational functional of the Generalized Gradient Approximation (GGA) was used.[25]

For the $Sr_3CrN_3H_x$ (0<x<3) calculations, an energy cut-off of 520 eV was employed. Any H atoms were added to the system as neutral species and all computations were spin polarized. Magnetic ions were initialised in a high-spin ferromagnetic configuration and then allowed to relax during each calculation. To sample **k**-space we used a 6 x 6 x 9 Monkhorst-Pack set of **k**-points with the tetrahedron method.[26] Electronic convergence criterion was set to $1 \times 10^{-6}$ eV and ionic convergence criterion was set to $1 \times 10^{-5}$ eV/Å in all cases. The $H_2$ gas was calculated in a unit cell of 10 Å$^3$ with a gamma centred single point **k**-grid.

To investigate potential diffusion pathways for mobile H- we performed a series of Climbing Image Nudged Elastic Band (CI-NEB) calculations with fixed mid-points, as implemented in VASP combined with the VTST-Tools[27] by Henkelman *et al*.[28]

Finally the screened–exchange hybrid density functional of Heyd-Scuseria-Ernzerhof (HSE06)[29] was used for calculations of electronic properties. This method is known to correct the underestimation of band gap and over delocalisation of conventional DFT.[30] For plotting the band structures we use Pymatgen's "electronic_structure" module BSPlotter.[31]

## Results and discussion

The crystal structure of $Sr_3CrN_3$ has the hexagonal space group P63/m, with lattice constants *a* = 7.84, *b* = 7.84, *c* = 5.24, α = 90.00, β = 90.00, γ = 120.00. The lattice constants obtained from the DFT structure relaxation compare well to the experimentally determined values (see **Table 1**).[32] The structure contains trigonal-planar [CrN$_3$]$^{5-}$ anions and Sr$^{2+}$ cations arranged to form 1D channels as shown in **Figure 1**. This 1D channel contains the excess electron density that allows the material to be defined as an electride.[33]

In $Sr_3CrN_3$, Sr and N ions occupy Wyckoff 6h sites and Cr occupies 2c sites. The remaining high-symmetry Wyckoff sites are vacant in the stoichiometric material and are therefore available as potential sites to accommodate anionic hydrogen.

To analyse the ability of $Sr_3CrN_3$ to accommodate intercalated hydrogen, we consider all possible H$^-$ configurations for *x*(H) = 1 to 6, where x(H) = 6 corresponds to all available Wyckoff sites being occupied by H. The intercalation energy as a function of x(H) is calculated as

$$\Delta E = E_{Sr_3CrN_3H_x(s)} - (E_{Sr_3CrN_3(s)} + E_{0.5xH_2(g)}) \qquad (1)$$

The energy change (ΔE) for all 25 total combinations of hydride position or listed in **Table S1** and plotted in **Figure 2,** although symmetry equivalence means fewer than 25 data points are visible in the plot. These results show that under standard conditions one intercalated hydride per formula unit has the lowest energy overall compared to the pure electride. There are two formula units of $Sr_3CrN_3$ in a unit cell, each one providing an excess electron that makes the material an electride. Thus, the addition of two hydrogen atoms to the unit cell creates the material $Sr_3CrN_3H$, which agrees with experimental reports on the related material $Ba_3CrN_3H$.[34] The positions of the hydride ions corresponding to the lowest energy are on the 2b Wyckoff positions with the fractional unit cell coordinates (0,0,0) and (0,0,0.5). The continued addition of hydrides is still favourable up to the addition of the 3rd hydride ion per formula unit, at which point the change in energy becomes positive. This indicates that increasing chemical potential of H would be needed for additional uptake of H to proceed.

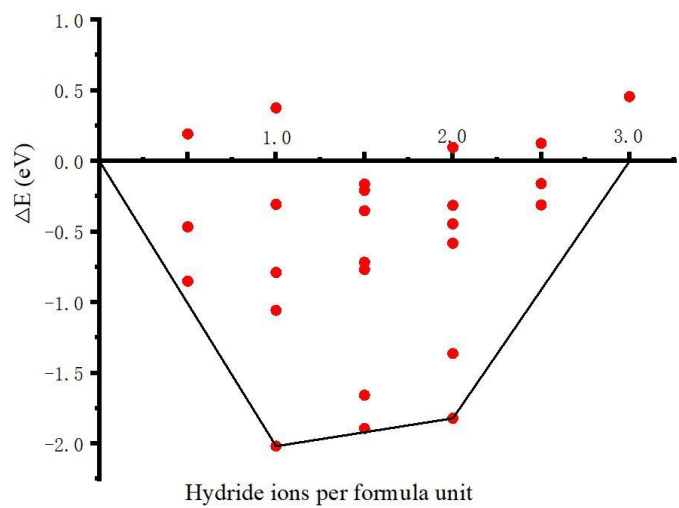

**Figure 2** The energy change as a function of hydrogen atoms added to $Sr_3CrN_3$ relative to the phase pure material. The lowest energies are connected to form a convex hull. The full data for this figure, corresponding to 25 total calculations, can be found in the Supporting Information.

McColm *et al*.[35,35] previously reported promise for electrides in the area of hydrogen storage. However, we find that under standard conditions at most 2.5 hydride ions per formula unit can be placed in the $Sr_3CrN_3$. Given that the volume of the unit cell is 282.68 Å$^3$ we calculate the corresponding volumetric $H_2$ density ($V(H_2)$) and relative gravimetric $H_2$ density ($\omega(H_2)$) as 29.61 kg/m$^3$ and 0.70 % respectively. These are calculated using Equations 2 and 3, below. Compared with other hydrogen storage materials, the hydrogen storage density of $Sr_3CrN_3$ is low—even lower than pressurised $H_2$ gas—owing to the composition and structure of our material.[37] Firstly, there are two rather heavy metal elements in this material, which makes the relative molecular mass large. Secondly, $Sr_3CrN_3$ has a 1D cavity channel, so the available volume is relatively low compared to other open-pore materials.

$$V(H_2) = \frac{m(H_2)}{v} \qquad (2)$$

$$\omega(H_2) = \frac{N\ (H_2) \times M\ (H_2)}{M\ (Sr_6Cr_2N_6H_5)} \times 100\% \qquad (3)$$

**Hydride migration pathways and diffusion coefficients**

We continue to assess the hydride ion migration properties of $Sr_3CrN_3$. Specifically, we calculate the energy barrier to ionic diffusion which directly relates to diffusivity and hence various device properties including necessary operating temperature and power output.[38] According to the results above, (0,0,0) and (0,0,0.5) are the optimum positions for hydride ions in the material. Therefore, these two positions are used as the initial and final positions for exploring the migration path of hydride ions using the CI-NEB method, as shown in **Figure 3**. These positions are within the one-dimensional cavity channel that is surrounded by strontium cations in the material.

The CI-NEB method generates a direct linear path connecting the start and end points with a very low energy barrier of 0.30 eV as shown in **Figure 4.** Even if the atoms of the host material remain fixed at the phase-pure positions, the energy is still found to be remarkably low at 0.35 eV. Our calculated energy barrier is smaller than those reported in the literature for the fastest reported hydride ion conducting materials: 0.52 eV for barium hydride,[12] or 1.2 eV for oxygen substituted lanthanum hydride.[10] Furthermore, proton migration is usually associated with an activation energy of higher than 0.5 eV in oxide materials,[39] despite protons having a smaller ionic radii than hydrides. As a result, these are among the most favourable properties for ionic diffusion reported previously.

We sample increasing displacement from the linear path within the channel to confirm the route calculated here is the ground state. We also analysed alternate migration routes outside of the one-dimensional channel. When we select (0.10,0.25,0.25) as midpoint, the migration barrier is 2.40 eV; for (0.25,0,0.25) the barrier is 2.64 eV and for (0.15,0.10,0.23), the barrier is 3.94 eV. Finally, we also sampled the paths along the other 2 unit-cell axes with *a* direction (0, 0, 0 -> 0, 0, 0.5) giving a barrier of 5.04 eV and *b* direction (0, 0, 0 -> 0, 0.5, 0) giving a barrier of 4.92 eV. Thus, we conclude that migration within the electride channel is highly favourable relative to alternate routes through the material. The large difference between energy barriers means that this material will be highly directional in its ionic migration, suggesting that so-called superionic fast conduction may be possible.[40]

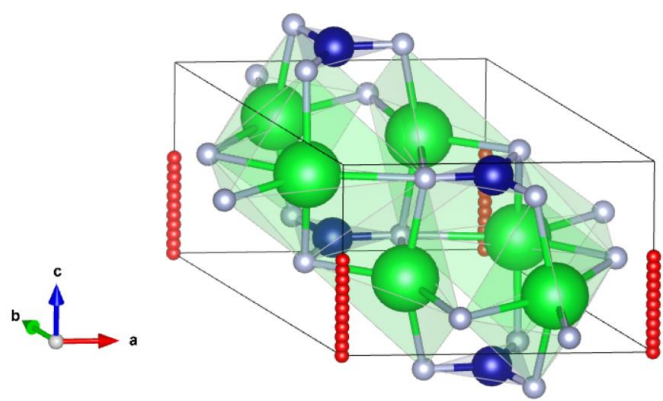

**Figure 3** The lowest energy hydride ion pathway in the $Sr_3CrN_3$ structure. The green spheres are Sr, the blue spheres are Cr, the light grey spheres are N, and the red spheres are H.

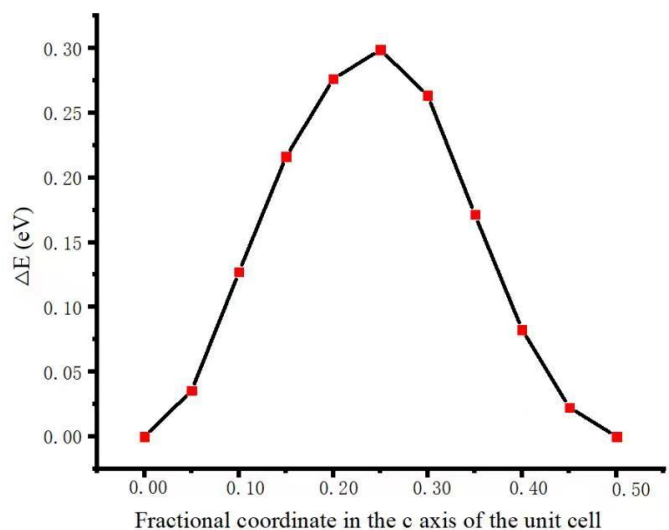

**Figure 4** The energy barriers from the CI-NEB method for the in-channel migration corresponding between (0,0,0) and (0,0,0.5) internal coordinates as shown in **Figure 3**.

Using the obtained potential energy barrier for diffusion, we estimate the diffusivity D using

$$D = D_0 e^{\frac{-Q}{K_B T}} \qquad (4)$$

Here, $Q$ is the energy barrier, $K$ is Boltzmann's constant, and $T$ is the temperature. $D_0$ is a constant pre-factor that we take from experimental reports of hydride ions diffusing through Fe metal, i.e. $D_0$ = 6.3×10$^{-3}$ cm$^2$/s. [41] The $Q$ value reported for this process is 0.44 eV, which is close to the value we find for $Sr_3CrN_3$ and in both cases the diffusing ion is a hydride. This calculation gives an approximate value of $D$ = 5.37×10$^{-8}$ cm$^2$/s for hydride in $Sr_3CrN_3$ at 298 K. Our result is similar to the value of $D$ = 1.4×10$^{-8}$ cm$^2$/s reported for fluoride ions in the electride $Y_2CF_2$,[42] Considering together, these results highlights that potential for electrides as general anion conductors.

**Electronic structures of the $Sr_3CrN_3$ and $Sr_3CrN_3H$**

We consider the effect of hydride intercalation into $Sr_3CrN_3$ on the electronic structure to illucidate the nature of the interaction between the hydride and the anionic electron and to ascribe possible application in devices. We use the hybrid functional HSE06 to calculate the electronic properties of $Sr_3CrN_3$ and $Sr_3CrN_3H$, based on occupation of 2H on the 2b site, i.e., the lowest energy configuration under standard conditions. The band structure diagrams for $Sr_3CrN_3$ and $Sr_3CrN_3H$ are shown in **Figure 5**. We find that $Sr_3CrN_3$ is metallic, which agrees with earlier analysis that places the anionic electron density of the electride at the Fermi level.[19] On the addition of the hydride ions, however, the compound becomes a wide band-gap semiconductor, presenting a band gap value of 3.0 eV. This again is consistent with analogous data reported for the related material $Ba_3CrN_3H$,[34] as well as for other electrides in the literature.[43] The large shift in electronic properties suggests that the anionic electrons reduce the hydride and are then not available in the structure. This is further corroborated by the fact that the bands of the hydride phase are relatively flat across the entire Brillouin zone, meaning that charges are tightly bound.

Druffel et al.[44] have suggested that electrides might find use in high-capacity electrodes that realise electron–anion reversible exchange at room temperature. In the case of $Sr_3CrN_3$ we predict no significant structural reorganisation and very small changes in unit cell volume upon hydrogen intercalation. Our analysis of the electronic nature of the material precludes such an application as the intercalated $Sr_3CrN_3H$ is strongly insulating, whereas an electrode is typically sought to be metallic in nature. What's more the flat bands mean a high charge effective mass and low conductivity, which is again unfavourable for application as an electrode. The hydride capacity is also relative low for such an application. However, the wide-band gap in the presence of hydride ions suggests that the material could be employed as an electrolyte in a fuel cell configuration, wherein a constant flow of hydride ions could maintain electrically insulating behaviour, which is what forces the electrons around the external circuit to produce electricity available for work. Finally, the band structure shows that the spin-up and spin-down states are degenerate, which indicates no intrinsic ferromagnetism for either the hydride or phase pure material.

## Conclusions

In summary, we have investigated the electride $Sr_3CrN_3$ as a hydrogen storage material and hydride ion conductor. The overall energy minimum under standard conditions is obtained on addition of one hydride per unit formula forming $Sr_3CrN_3H$, although up to 2.5 should still be thermodynamically stable, forming $Sr_6Cr_2N_6H_5$. While uptake and desorption of hydrogen is readily observed in experiment for electrides,[17,18,15] this material is not likely to be a candidate of a hydrogen storage material due to the heavy constituent elements and relatively unopen structure.

On the other hand, we find $Sr_3CrN_3H$ to be an excellent conductor of hydride ions. Within the 1D channel hydrogen ions show a migration barrier of 0.30 eV, which is significantly lower than for even some of the best performing hydride-ion conductors reported to date, e.g., 0.52 eV for barium hydride,[12] or 1.2-1.3 eV for oxygen substituted lanthanum hydride.[10] While it is tempting to attribute the low diffusion barrier to the anionic electrons in a one-dimensional channel in the material facilitating conduction, the electronic structure indicates that the excess electrons are used in reducing the H to H$^-$; becoming a wide-band gap ionic material in the process. Thus, we foresee ample scope for future study of ionic conduction in this material and other electrides in the future. Overall, we find that $Sr_3CrN_3H$ exhibits exceptional ionic transport properties and could be particularly useful for energy storage devices, catalysis or gas separation applications.

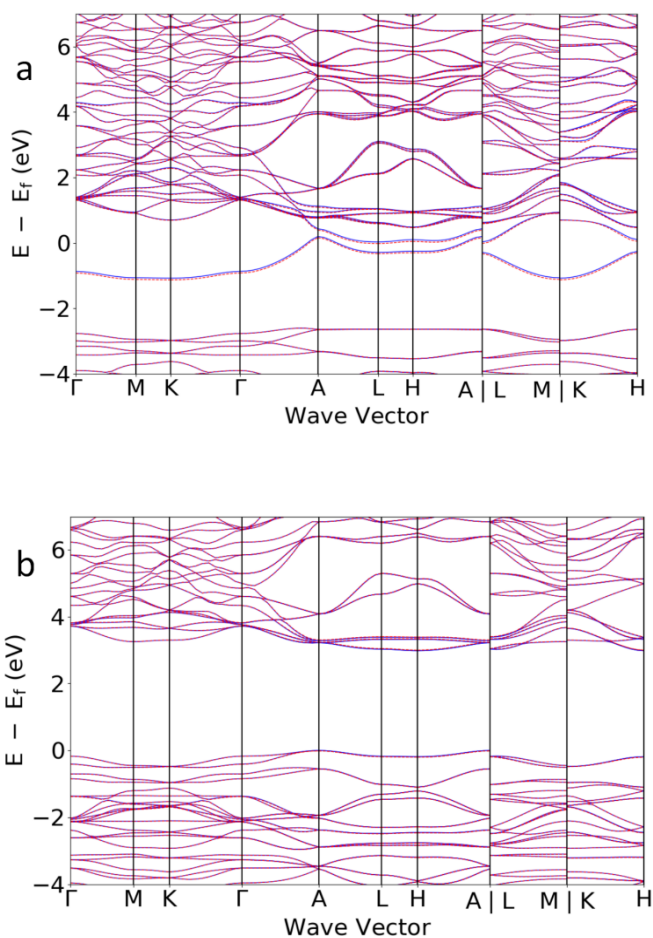

**Figure 5** (a) The band structure of $Sr_3CrN_3$ showing metallic behaviour and (b) the band structure of $Sr_3CrN_3H$ showing a relatively wide band gap. The red and blue states are spin up and spin down respectively.

## Conflicts of interest

There are no conflicts to declare

## Acknowledgements

The L.A.B acknowledges support by the Shanghai Municipal Science and Technology Commission Program, number 19010500500, and the Natural National Science Foundation of China (NSFC) number 51950410585.